\documentclass[12pt]{article}
\usepackage{graphicx}
\newlength{\szovszel}\newlength{\perwidth}

\newcommand{\eqref}[1]{(\ref{#1})} \renewcommand{\c}[1]{{\cal #1}}
\renewcommand{\d}{\partial} \newcommand{\nn}{\nonumber\\}
 \newcommand{\rh}{\varrho}
\newcommand{\exv}[1]{\left\langle{#1}\right\rangle}

\newcommand{\ep}{\varepsilon} \newcommand{\x}{{\bf x}}
 \newcommand{\q}{{\bf q}}
\renewcommand{\k}{{\bf k}} \newcommand{\p}{{\bf p}}
\newcommand{\T}{\textrm{T}} \newcommand{\Tr}{\mathop{\textrm{Tr}}}
\renewcommand{\Im}{\,\textrm{Im}\,}

\newcommand{\pint}[2]{{\int\!\frac{d^{#1}#2}{(2\pi)^#1}\,}}
 \newcommand{\cP}{\mathop{\cal P}}
\newcommand{\dlr}{\mathop{\mbox{$\d$\hspace*{-0.6em}\raisebox{0.8em}{$\scriptstyle\leftrightarrow$}
      \hspace*{-0.4em}}}}

\begin{document}
\pagestyle{empty}
\begin{flushright}
  CERN-TH/2001-361\\
  hep-ph/0112188
\end{flushright}
\vspace*{5mm}
\begin{center}
  {\bf TIME EVOLUTION IN LINEAR RESPONSE: BOLTZMANN EQUATIONS AND BEYOND}
  \vspace*{0.8cm}
  
  A. Jakov\'ac\footnote{e-mail: Antal.Jakovac@cern.ch}\\
  
  \vspace{0.3cm} {\em Theory Division, CERN, CH-1211 Geneva 23,
    Switzerland}
  
  \vspace*{2cm} {\bf ABSTRACT}
\end{center} 
\vspace*{5mm}

\noindent  
In this work a perturbative linear response analysis is performed for
the time evolution of the ``quasi-conserved'' charge of a scalar
field. One can find two regimes, one follows exponential damping,
where the damping rate is shown to come from quantum Boltzmann
equations. The other regime (coming from multiparticle cuts and
products of them) decays as power law. The most important,
non-oscillating contribution in our model comes from a 4-particle
intermediate state and decays as $1/t^3$. These results may have
relevance for instance in the context of lepton number violation in
the Early Universe.

\vspace*{0.5cm}
\begin{flushleft} CERN-TH/2001-361\\ December 2001
\end{flushleft}
\vfill\eject

\setcounter{page}{1} \pagestyle{plain}

\section{Introduction}

Recently, in a series of papers \cite{JMY}, Joichi, Matsumoto and
Yoshimura analyzed the effects of the quantum-improved Boltzmann
equations. The authors developed a model of unstable particles and
found that their Boltzmann equations generate a
non-exponentially-suppressed equilibrium distribution function for
heavy particles at low temperatures. This deviation from the standard
picture was due to cut contributions far from the quasi-particle pole.
These works were later criticized by several authors \cite{Sred,
  BraaJia, BuPie}, their main point being that the definition of
particle number is delicate for unstable particles. Redefining
particle number may change the coefficient of the non-exponential term
\cite{BraaJia}, or completely cancel it \cite{Sred}. In \cite{BuPie}
the authors argued that the naive definition needs renormalization in
perturbation theory and proposed a non-perturbative definition for the
number density.

Non-quasi-particle properties of the quantum fields may change also the
time evolution of the physical observables, as was also emphasized
in \cite{JMY}. The same phenomenon was found also in the time
evolution of bosonic condensates \cite{Boyan,JPSZ}. In these latter
cases power-law decay was shown up even in the linear response theory.

In this paper we use linear response perturbation theory to
investigate the long-time behavior of composite operators. As a
specific example we consider the ``quasi-conserved'' charge of a
scalar field. In this way we cannot address the equilibrium state;
still, it is a physical scheme at which practically all systems must
arrive after long time evolution. The numerical advantage is that all
expectation values are defined in equilibrium, so that they can, in
principle, be calculated without additional approximation. It is very
important that the different contributions are additive, which makes
it possible to identify different physical effects without entangling
them.

In fact there is a close analogy in the formalism describing the time
evolution of the field condensates \cite{Boyan,JPSZ} and the scalar
charge operators; the details will be explained in the paper. In
linear response theory the field condensates are proportional to the
retarded Green's function; in the case of the charge $Q$ we have
instead $QF$ retarded Green's function (with arbitrary $F$). The
discontinuity of the retarded propagator for free case is a delta
function concentrated on the mass shell. In the free theory $Q$ is
conserved, thus the discontinuity of $QF$ retarded Green's function is
$\sim\delta(k_0)$ -- this is the ``mass shell'' of the $QF$ spectral
function. In the interacting case, at small coupling, the mass shell
still represents quasi-particles. In a similar way the charge,
although not conserved, characterizes the system well. In analogy to
quasi-particles we call it quasi-conserved charge.

The shift of the mass shell is not a perturbative effect. Indeed, when
we calculate the propagator in perturbation theory we find IR
divergences on the mass shell. One uses Schwinger--Dyson equations to
resum these divergences, and a Breit--Wigner approximation to describe
the shift of the mass shell. The same phenomenon appears also in the
$QF$ propagator: the shift of the ``mass shell'' causes IR divergences
in perturbation theory at $k_0=0$.  In coordinate space these
divergences are manifested as secular terms.  These, as was shown by
\cite{Boyan-DRG}, can be resummed to quantum Boltzmann equations.  In
momentum space the same phenomenon shows up as pinch singularities
\cite{LandvanWeert,Altherr,CarrDefuThoma, GreiLeu}. In order to resum
them we have to consider ladder diagrams \cite{Mahan,Langer,
  Carrington,Jeon,AMY}, which we will show to result in Kadanoff--Baym
equations. These are, at lowest order \cite{BuchFred}, the usual
Boltzmann equations. Thus the analogy of the Schwinger--Dyson
resummation in Breit--Wigner approximation is the use of Boltzmann
equations in the composite operator case.

The analogy, however, extends even more. In the propagator we find,
apart from the broadened mass shell, other structures as well. Most
important are the analytic defects (e.g. multiparticle cuts), which
play an important role in the long-time evolution, as they give
power-law decay \cite{JMY,Boyan,JPSZ}. These are not related to the
broadening, and they are perturbative in the massive case. We find
similar defects in the $QF$ propagator, giving power-law decay at
large times, which are independent of the Boltzmann equations.

This paper is structured as follows. In Section~\ref{sec:longtime} we
work out how initial conditions can be taken into account in linear
response theory and what kind of long-time behavior can be expected.
It has two ingredients: a pole contribution, which yields exponential
damping, and a power-law correction coming from other analytic defects
(such as cuts). In Section~\ref{sec:dampingrate} the first
contribution is considered and it is shown that this comes from the
quantum Boltzmann equations (Kadanoff--Baym equations). In
Section~\ref{sec:cut} the off-shell analytic defects are investigated.
Multiparticle cuts give a power-law, oscillating time dependence. It
is shown, however, that they can be combined even at the linear
response level giving non-oscillating time dependence. In
Section~\ref{sec:conclusion} we summarize our findings.

\section{Long-time behavior in linear response theory}
\label{sec:longtime}

\subsection{The model}

Inspired by the discussion about the particle number, we choose a
well-measurable observable: the charge of a bosonic field
(cf. \cite{Sred}). The Lagrangian
\begin{equation}
  \c L = -\Psi^\dagger (\d^2+m_\Psi^2)\Psi - \chi^\dagger
  (\d^2+m_\chi^2)\chi - \frac12 \Phi (\d^2+m_\Phi^2)\Phi -
  h\Phi(\Psi^\dagger\chi+ \chi^\dagger\Psi)
  \label{Lag}
\end{equation}
contains a heavy charged $\Psi$ particle (with mass $m_\Psi$) and a
light charged $\chi$ particle (with mass $m_\chi$) coupled through a
Yukawa term with a scalar $\Phi$ (with mass $m_\Phi$).

Without the Yukawa term, the U(1) phase symmetry would be perfect, and
the currents
\begin{equation}
  J^\Psi_\mu = \Psi^\dagger\,i\dlr\Psi\quad\textrm{and}\quad
  J^\chi_\mu = \chi^\dagger\,i\dlr\chi
\end{equation}
would be conserved quantities. For $h\neq0$ they are not conserved
\begin{equation}
  \d J^\Psi = ih\Phi(\chi^\dagger\Psi - \Psi^\dagger\chi), \qquad
  \d J^\chi = ih\Phi(\Psi^\dagger\chi - \chi^\dagger\Psi),
\end{equation}
only their sum is. Our final goal is to describe the time evolution of
the charge of $\Psi$ and $\chi$
\begin{equation}
  Q^\Psi(t) =\exv{\hat Q^\Psi(t)} = \Tr \hat\rho Q^\Psi(t),\qquad \hat
  Q^\Psi(t)= \int\! d^3\x\,J^\Psi_0(\x,t),
\end{equation}
and similarly for $\chi$, where $\hat\rho$ is the initial density
matrix. In order to avoid complications with the chemical potential,
we assume that the total charge is zero.

\subsection{Linear response theory}

Tracing the time evolution, starting from a non-equilibrium density
matrix, is usually a hard task \cite{directrho}. Here we follow a way
where we construct the initial state from equilibrium by acting on the
system with a time-dependent external force -- after all this is the
way one acts in a real experiment. At $t=-\infty$ we start from
equilibrium, then we prepare the initial conditions by changing $H\to
H+H_{ext}(t)$ for times $-\infty<t<0$, finally we turn off the
external force and start to measure for $0<t<\infty$.

This picture can be formulated in path integral representation, since
at $t=-\infty$ we have equilibrium ensemble. For an observable
$O(x)$
\begin{equation}
  \exv{O(x)} = \frac1Z \int\limits_{KMS}\!\!O(x)\, e^{iS -i
  \int_c \zeta F},
\end{equation}
where $\int_{KMS}$ denotes path integration where the paths are
subject to KMS conditions along the applied Keldysh contour $c$. The
external force is $F$ and its explicitly time-dependent amplitude is
denoted by $\zeta(t)$ (here $t$ is the real time, not the contour time
variable).

By the end of a realistic time evolution, we deviate only slightly
from equilibrium; therefore we expect that the corresponding state can
be produced using a weak force. That means that in this case we can
use linear response theory. If in equilibrium $\exv{O}_{eq}=0$ then we
have
\begin{equation}
  \exv{O(x)} = -i\int d^4x'\,\Theta(t-t') \exv{[O(x),F(x')]}_{eq}
  \zeta(x').
\end{equation}
Here the expectation value of the commutator has to be taken in
equilibrium, and therefore it is a function of $x-x'$. We denote
\begin{equation}
  i\c G_R^{(OF)}(x) = \Theta(t) \exv{[O(x), F(0)]}_{eq},\qquad
  \rh^{(OF)}(x) = \exv{[O(x),F(0)]}_{eq};
\end{equation}
note that in Fourier space $\rh^{(OF)}(k) =
i\mathop{\textrm{Disc}}\limits_{k_0} \c G_R^{(OF)}(k)$.

For our concrete observable,
\begin{equation}
  Q(k_0) = \exv{J_0(k_0,\k=0)} = \c G_R^{(J_0F)}(k) \zeta(k)\bigr|_{\k=0}.
\label{QGRZ}
\end{equation}
This is very similar, in spirit, to the usual Kubo formulae
\cite{LeBellac}, the only difference being that now the external
force is not fixed. In principle we can imagine a lot of possible
choices (e.g.  a plausible one $F\sim H_I$ where $H_I$ is the
interaction Hamiltonian); what is only important is that it should
create a net $Q(0)$, i.e. $[\hat Q,F]\neq0$.

Having the formal result in hand, there is still a question of how to
characterize the process and how to compute the specific numbers. In
general we expect that, close to $t=0$ when the external force was
switched off, the time evolution contains a lot of transients. This is
reflected also in the fact that $Q(t)$ depends strongly on the choice
of $F$. Only after a long time do we expect that the time dependence
is independent of the initial conditions. Concentrating on this region
we now examine the long-time behavior of an expression like
\eqref{QGRZ}.

\subsection{Long-time behavior: the general case}

\begin{figure}[htbp]
  \begin{center}
    \includegraphics[height=3cm]{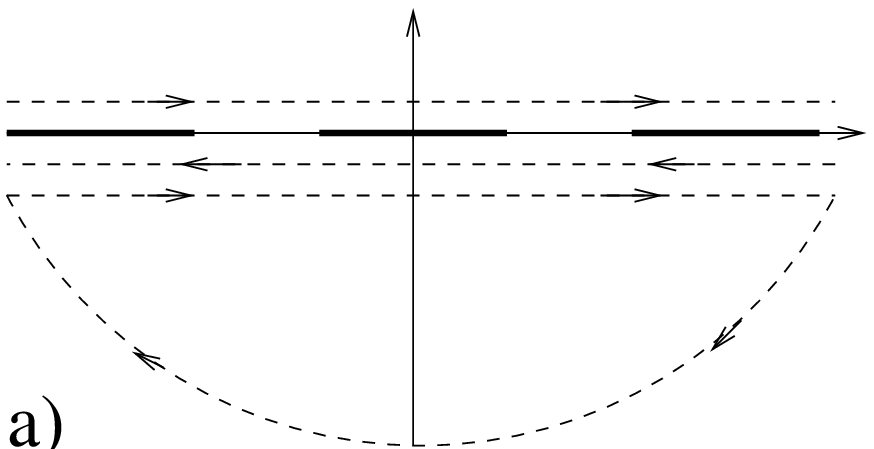} \hspace*{2cm}
    \includegraphics[height=3cm]{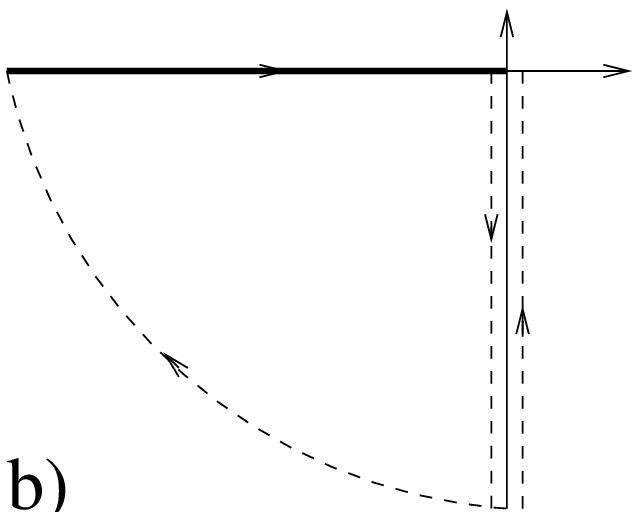}
    \caption{Transformation of the integration contours.}
    \label{fig:intcont}
  \end{center}
\end{figure}

This analysis follows closely the one of Refs.~\cite{Boyan,JPSZ},
including also the $\zeta$ dependence. We examine the long-time
behavior of
\begin{equation}
  f(t)= \int\!\frac{dk_0}{2\pi}\, e^{-ik_0 t} \,\c G_R(k_0)\zeta(k_0),
\end{equation}
where $\c G_R$ is a general retarded Green's function, $\zeta$ is the
amplitude of the external force. We assume that the analytic defects
of $\zeta(k_0)$ are on the upper half-plane and the ones of $G_R$ are
along the real axis. Then, taking into account that using retarded
Green's functions means that we have to use a contour that runs
slightly above the real axis in $k_0$, we can extend the integration
region for $t>0$ as is shown in Fig.~\ref{fig:intcont}a (the
semi-circle at infinity gives zero contribution).
Under our assumptions the closed circle on the lower half-plane gives
zero, while the rest picks up the discontinuity of $G_R$
\begin{equation}
  f(t) = \int\!\frac{dk_0}{2\pi i}\, e^{-ik_0 t} \,\rh(k_0)
  \zeta(k_0).
\end{equation}
The spectral function $\rh$ may contain poles or discontinuities in
derivatives of some order (ie. thresholds), but these defects form a
discrete set. In between, the spectral function is analytic. We
transform the piecewise analytic parts further. For a single interval
running from $-\infty\to a$:
\begin{equation}
  \int\limits_{-\infty}^a\frac{dk_0}{2\pi i}\, e^{-ik_0 t} \,
  \rh(k_0) \zeta(k_0) = e^{-iat} \int\limits_{-\infty}^0
  \frac{dk_0}{2\pi i}\, e^{-ik_0 t} \, \rh(k_0+a) \zeta(k_0+a).
\end{equation}
Now we add the zero contribution of a quarter circle at infinity and a
contour running up and down on the negative imaginary axis as shown by
Fig.~\ref{fig:intcont}b. The closed loop picks up the contributions
of the poles inside; in the rest we change variables $ik_0\to y$ and
write
\begin{equation}
  \mathop{\textrm{Res}}\limits_{\omega\in\textrm{\scriptsize loop}}
  \left[\rh(\omega)\zeta(\omega) e^{-i\omega t}\right] \,+\,
  e^{-iat} \int\limits_0^\infty \frac{dy}{2\pi}\, e^{-y t} \,
  \rh(a-iy) \zeta(a-iy).
\end{equation}
The poles give exponential damping, while the rest, in general, can be
an arbitrary function.

So far we were exact, now we make an approximation valid for long
times: we power-expand $\rh$ and $\zeta$ around $y=0$. Since for
$\zeta$ the threshold $a$ is not a special point we expect that it
starts with a constant. On the other hand $\rh$ usually starts with
$Ry^\alpha$, where $R$ is a constant and $\alpha\neq0$. The omitted
terms are at least one power of $y$ larger than the leading ones.
Therefore
\begin{equation}
  \int\limits_0^\infty \frac{dy}{2\pi}\, e^{-y t} \, \rh(a-iy)
  \zeta(a-iy) \stackrel{t\to\infty}{\longrightarrow}
  \int\limits_0^\infty \frac{dy}{2\pi}\, e^{-y t} \, Ry^\alpha
  \zeta(a) \left(1+{\cal O}(y)\right) =
  \frac{R\zeta(a)\Gamma(\alpha+1)}{2\pi t^{\alpha+1}} \left(1+{\cal
  O}(\frac1t)\right).
\end{equation}
The most important term, therefore, comes from the leading power
behavior at the threshold. Finally we find for the long-time
behavior:
\begin{equation}
  f(t) \stackrel{t\to\infty}{\longrightarrow}
  \sum\limits_{\omega\in\textrm{\scriptsize poles}}\!\!
  \mathop{\textrm{Res}} \left[\rh(\omega)\zeta(\omega) e^{-i\omega
  t}\right] \, + \!\!\!\!
  \sum\limits_{a\in\textrm{\scriptsize thresholds}} \!\!\!\! (\pm)
  \frac{R\zeta(a)\Gamma(\alpha+1)}{2\pi}\,
  \frac{e^{-iat}}{t^{\alpha+1}},
\label{Qt}
\end{equation}
where the $+$ sign is valid for the starting and the $-$ is for the
ending of an interval.

An important feature of this formula is that the different terms have
independent, initial-condition-dependent weights (values of $\zeta$ at
different points are independent). Constraints can be formulated in
the form of sum rules, as for example
\begin{equation}
  Q(t=0) = \int\!\frac{dk_0}{2\pi i}\, \rh(k_0) \zeta(k_0),
\end{equation}
but it gives information on the integral properties, and not for
single values which appear in \eqref{Qt}. So, practically, the
relative weights of the different parts in the long-time expression
\eqref{Qt} are arbitrary.

This behaviour is, in fact, well known if we have a finite number of
poles only,
\begin{equation}
  f(t) = \sum\limits_{i=0}^{N-1} c_i e^{-i\omega_i t},
\end{equation}
which can be the result of an $N$th order ordinary linear differential
equation with constant coefficients. The smallest $\Im\omega_i$ will
be dominant for long times, the other frequencies represent a
transient behaviour. The coefficients $c_i$ are free parameters coming
from the initial conditions. For a finite number of poles, sum rules
of the form
\begin{equation}
  f^{(n)}(t=0) = \sum c_i (-i\omega_i)^n,\qquad n=0\dots N-1
\end{equation}
(the superscript $(n)$ stands for the $n$th derivative) can fully
determine the coefficients -- ie. the knowledge of $N$ derivatives at
the initial time is needed. In case of continuous spectral functions,
however, detailed knowledge of the history is necessary for their
complete determination.

We can also find the analogue of the power law behaviour for the
finite case. If we consider two oscillators working at nearby
frequencies $\omega\pm\Delta$, then
\begin{equation}
  \cos(\omega+\Delta)t + \cos(\omega-\Delta)t = 2\cos\omega t \cos
  \Delta t,
\end{equation}
the result is an average frequency oscillation modulated by a low
frequency oscillation. In this way high frequency modes can have IR
behaviour. When the number of the oscillators grows and at the same
time the frequency difference vanishes we have
\begin{equation}
  \frac1N \sum\limits_{n=0}^{N-1} e^{i(\omega+n c/N)t} =
  \frac{e^{i\omega t}}N \frac{1-e^{ict}}{1-e^{ict/N}} \stackrel{1\ll
  ct\ll N}{\longrightarrow} \frac{ie^{i\omega t}}{ct},
\end{equation}
i.e. we arrive indeed at a power-law decay. For $N\to\infty$ this will
be valid for all times.

\subsection{Long-time behaviour for quasi-conserved quantities}

\begin{figure}[htbp]
  \begin{center}
    \includegraphics[height=4cm]{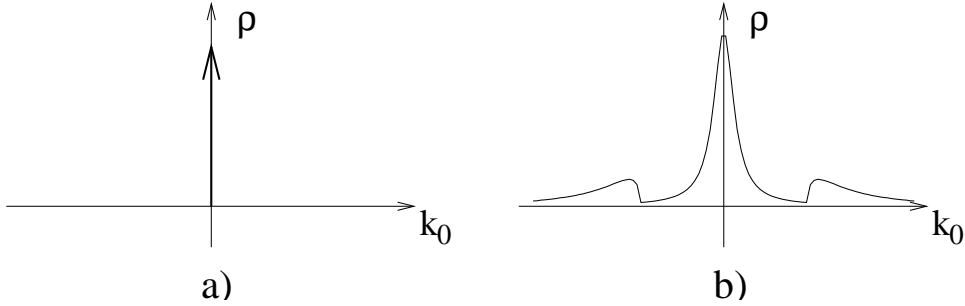}
    \caption{Schematic plot of the spectral function for {\bf a.)}
      conserved quantity and {\bf b.)} weak breaking.}
    \label{fig:rhQ}
  \end{center}
\end{figure}

The structure of $\rh$ is special for an observable that is conserved
in some limit, and its breaking is weak. In our specific case in the
$h\to 0$ limit $Q$ is conserved, so we expect that for small Yukawa
couplings the following will be true. If there were no breaking at
all then $\dot Q = i[H,Q]=0$, therefore
\begin{equation}
  \rh(t,\k=0) = \exv{[\hat Q(t),F(0)]}_{eq} = \exv{[\hat
  Q(0),F(0)]}_{eq} = \rh(0,\k=0).
\end{equation}
Its Fourier transform therefore is $\rh(k_0)\sim \delta(k_0)$, as can
be seen in Fig.~\ref{fig:rhQ}a.
If $Q$ is not conserved, we expect, in general, that we have non-zero
$\rh$ also for other values of $k_0$. Close to the conserved limit, we
shall recover something like Fig~\ref{fig:rhQ}b. The delta peak at
zero $k_0$ will be broadened, but it still represents a pole near the
origin. Moreover we expect to find some other structures (analytic
defects) as well. From this spectral function, according to the
previous subsection, we have the time dependence for long times
\begin{equation}
  Q(t) = Q_1 e^{-\Gamma t} + Q_2 \frac{\cos (\Omega t
    +\phi)}{t^{\alpha +1}} + \dots,
  \label{Qbehav}
\end{equation}
where $\Gamma$ is the shift of the $k=0$ pole, $\Omega$ is the
position of the threshold with the smallest $\alpha$. The coefficients
$Q_1,\, Q_2$ and $\phi$ depend on the initial conditions.

One could think that power-law terms dominate for really long times.
This is true; still, there are effects which make them less relevant.
One is that usually $\Omega\sim m$ (a mass scale in the system), and
so they are fast oscillating functions. Thus they can hardly influence
physics on scales larger than $m^{-1}$. The $\Omega\sim0$ case is
exceptional. This, however, shows up at higher order in perturbation
theory (or comes from non-linear effects -- see later) and $Q_2$ is
proportional to some powers of the coupling constant $h$.  Thus
$Q_1\gg Q_2$ (unless we have very special initial conditions), and the
first term will be larger for a large period of time.  Analyzing the
functions we find that the two terms are equal at time $t$, where
\begin{equation}
  \tau < \Gamma t < 1.582 \tau,\qquad \tau = 
  \ln\left[\frac{Q_1}{Q_2}\left(\frac{\alpha+1}\Gamma\right)^{\alpha+1}
  \right].
\end{equation}
Since $Q_1/Q_2$ and $\Gamma$ are polynomials of $h$, the first term is
the leading one up to times $\Gamma t \sim - C \ln h$, where
$C\sim5$--$10$ typically (for example if $Q_2/Q_1\sim h^2$, $\alpha+1=3$
and $\Gamma\sim h^2$ then $C=8$). During this time the charge drops a
factor of $e^{-\Gamma t} \sim h^C$, i.e. for small couplings the power
law regime may be unobservable. A hint for the presence of such a
regime, however, can be seen in numerical simulations \cite{BorsSzep}.

On the other hand, if $h$ is not small, then the exponential behaviour
may be suppressed. In fact, only in quasi-particle picture, when
$\Gamma\ll$ other mass scales, do we expect exponential damping.

\section{The damping rate}
\label{sec:dampingrate}

In the previous section we outlined the general long-time behaviour
for a quasi-conserved quantity such as the particle charge. The two
relevant effects found are broadening of the delta peak at $k_0=0$,
and analytic defects at $k_0=0$. In this section we make some
remarks on the broadening effect.

In our specific example at $h=0$ the charge is conserved, and we
assume that for small $h$ it still represents a pole on the second
Riemann sheet close to the origin. For small $k_0$ we thus expect that
the time-ordered or retarded propagator (at least the pole
contribution to that) can be approximated as
\begin{equation}
  G^{(QF)}(k_0) = \exv{\T\, \hat Q\,F}(k_0)\to  \frac C{k_0 +
  i\Gamma},\qquad \textrm{for}\quad k_0\to 0,\;\Im k_0<0.
\label{Glongtime}
\end{equation}
This form suggests that the broadening effect must show up as IR
divergences in the perturbation theory. Here, namely, we can calculate
only $h$-power corrections to the leading result, therefore the
appearance of $\Gamma$ in the denominator means that in perturbation
theory we find
\begin{equation}
  G^{(QF)}(k_0\to 0) \to \frac C{k_0} \sum\limits_{n=0}^\infty
  \left(\frac{-i\Gamma}{k_0} \right)^n.
  \label{expandform}
\end{equation}
The signal of this behaviour is the $\sim 1/k_0^n$ divergence of the
perturbative expression for small (but not zero!) $k_0$. 

Working in coordinate space the $k_0^{-n}$ divergences turn into
$t^{n-1}$ time dependence. These are the secular terms that appear in
all orders in perturbation theory, and that can be resummed according
to \cite{Boyan-DRG} as the quantum Boltzmann equation.

\begin{figure}[htbp]
  \begin{center}
    \includegraphics[height=3cm]{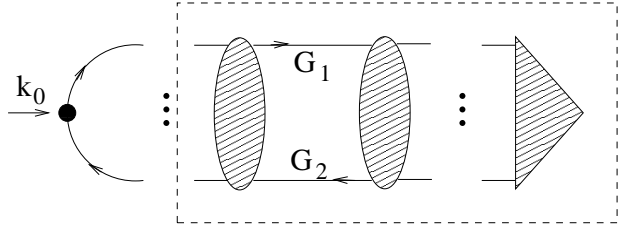}
    \caption{IR-divergent (ladder) diagram. The shaded insertions can
      represent rungs or self-energy insertions.}
    \label{fig:fermdiag}
  \end{center}
\end{figure}

Using perturbative methods in Fourier space, the origin of these IR
divergences is the pinch singularities \cite{LandvanWeert,Altherr}. To
see this consider a diagram contributing to the $QF$ retarded Green's
function with a 2-particle intermediate state, as can be seen in
Fig.~\ref{fig:fermdiag}.
The momenta of the propagators $G_1$ and $G_2$, because of
energy--momentum conservation, are $q+k_0$ and $q$, respectively. If
these propagators are retarded and advanced with the same mass, then
for $k_0=0$ the poles of the two propagators approach the real axis
from above and from below, thus ``pinching'' it, giving the divergent
result $G_R(q)G_A(q) \sim (\delta(q^2-m^2))^2$
\cite{LandvanWeert,Altherr}.  For finite $k_0$ we do not expect
divergences, but the $k_0\to 0$ limit must be infinite. In
Appendix~\ref{sec:pinch} we indeed find that, at the $k_0\to 0$ limit,
we have a singular behaviour for the products
\begin{equation}
  G_R\left(q+\frac k2\right)G_A\left(q-\frac k2\right)\to i\rh(q)
  \frac{d(q)}{2q_0k_0},\qquad G_{R/A}\left(q+\frac
  k2\right)G_{rr}\left(q - \frac k2\right) \to
  \frac{d(q)}{2q_0k_0}G_{rr}(q),
  \label{k0rule}
\end{equation}
and the inclusion of $G_R(q+k/2)$ yields an extra $d(q)/2q_0k_0$
factor, while the inclusion of additional $G_A(q-k/2)$ yields
$-d(q)/2q_0k_0$.  Therefore a ladder containing $n$ rungs or
self-energy insertions may yield a $k_0^{-n}$ IR divergence,
corresponding exactly to our expectations.

We can also argue that more particle intermediate states do not give
pinch singularities in the massive case. Consider a 3-particle
intermediate state with momenta $p,\,q$ and $-p-q$ ($k=0$ case).  We
expect an IR enhancement if the sum of two momenta is almost zero; let
us take $q\approx p+q$, i.e. $p\approx 0$. Then the propagator with
momentum $p$ gives $1/m^2$, where its mass is $m$, the other two
propagators are replaced using Appendix~\ref{sec:pinch}. One finds,
concentrating on this region,
\begin{equation}
  \sim d^4p\,d^4q\, \frac{\rh(q)}{m^2}\, \frac1{2 p_0 q_0 -
  \omega^2_{p+q} +\omega^2_q}.
\end{equation}
This expression is IR-safe, as can be seen for example by integrating
over the IR region $|p_0|,|\p|<\ep$. Therefore we cannot have any
$1/k_0$ enhancement from this insertion. This argumentation can be
extended to more internal lines.

There exists another possibility when there is an IR sub-divergence,
coming from a self-energy insertion on an internal line.  Since,
however, internal lines are equilibrium propagators, these pathologies
must finally cancel \cite{LandvanWeert}.

To resum the ladder diagrams we perform point splitting in a symmetric
way:
\begin{equation}
  \exv{J_\mu(x)} = \lim\limits_{y\to x} (i\d^x_\mu-i\d^y_\mu)
  \frac12 \exv{ \Psi(x)\Psi^\dagger(y) + \Psi^\dagger(y)\Psi(x)} =
  \lim\limits_{y\to x} (i\d^x_\mu-i\d^y_\mu) i\c{G}_{rr}^\Psi(x,y),
  \label{pointsplit}
\end{equation}
and similarly for $\chi$, where we used R/A formalism to define the
propagator \cite{RAformalism}. The $\c G$ propagator corresponds to
the boxed part in the diagram of Fig.~\ref{fig:fermdiag}.  In Fourier
space
\begin{equation}
  \exv{J_\mu(k)} = \pint4q (2q+k)_\mu i\c G_{rr}(q+k,q).
\end{equation}
Introducing Wigner transformation
\begin{equation}
  \c W(q,X) = \int d^4u \, e^{-iqu} i\c G(X+\frac u2,X-\frac u2),\qquad
  \c W(q,k) = i\c G(q+\frac k2, q-\frac k2)
  \label{Wignerdef}
\end{equation}
we can rewrite it as
\begin{equation}
  \exv{J_\mu(k)} = \pint4q 2q_\mu \c W_{rr}^\Psi(q,k).
  \label{JandG}
\end{equation}

A ladder diagram is a recursive structure, thus there must be a kernel
which relates a ladder containing $N+1$ and $N$ insertions. Keeping
only the pinch singular terms we can write $i\c G^{N+1}(p,p') =
1/{2k_0q_0} (K*i\c G^N)(p,p')$, where $q=(p+p')/2$ and $k=p-p'$ (cf.
\eqref{k0rule}). The sum for all ladders $\c G= \sum_N\c G^N$ then
satisfies $i\c G(p,p') = 1/{2k_0q_0} (K*i\c G)(p,p')$.  With the Wigner
transformed functions \eqref{Wignerdef}, this can be written as
$\c W(q,k) = 1/{2k_0q_0} (K*\c W)(q,k)$, or, in real space (performing
a Fourier transformation with respect to $k$),
\begin{equation}
  2i q_0 \d_t \c W(q,t) = (K*\c W)(q,t),
  \label{leq1}
\end{equation}
a first order differential equation.

This differential equation is unique up to leading order in the $k$
expansion. Let us assume, namely, that there is another equation
\begin{equation}
  2i q_0 \d_t \c W(q,t) = (K'*\c W)(q,t)
  \label{KadBayeq}
\end{equation}
satisfied by $\c W$. Then inserting the solution of \eqref{leq1} into
this equation we find
\begin{equation}
  (K-K')*\c W =0.
\end{equation}
This is an identity that must be fulfilled for $K$ and $K'$. This
shows that $K=K'$ on the space of the solutions of \eqref{leq1}.

Using this property we can use differential equations of the form of
\eqref{leq1} with different origin in order to resum pinch
singularities. And exactly of this form are the Kadanoff--Baym
equations \cite{KadBaym} in linear response. For the free theory,
namely, the propagators satisfy, because of the Schwinger--Dyson
equations,
\begin{equation}
  (\d_x^2-\d_y^2) \c G(x,y) =0 \quad\Rightarrow\quad 2iq_0 \d_t \c
  W(q,t)=0.
\end{equation}
In the case of interactions the right-hand side depends on $\c G$ (or
$\c W$), but in the linear response theory it is linearized and has
the form $K * \c W$ with some kernel. We therefore recover
\eqref{KadBayeq}. The uniqueness of this equation then implies that
the Kadanoff--Baym equations represent an adequate tool to resum the
pinch singularities of ladder diagrams in linear response theory.

So finally both the coordinate space method \cite{Boyan-DRG} and the
momentum space method outlined above support the statement that the
resummation of $1/k_0^n$ IR singularities of the perturbation theory
yields quantum Boltzmann equations (Kadanoff--Baym equations).
Therefore $\Gamma$ appearing in \eqref{Glongtime} must be calculated
using Boltzmann equations.

In our model to lowest order we have to take into account the decay
processes, which have a matrix element $h$. Therefore the Boltzmann
equations are
\begin{equation}
  2q\d_X N^\Psi(q,X) = h^2 \pint4\ell \rh^\chi(\ell) \rh^\Phi(q-\ell)
    \left[ (N^\Psi+1) N^\chi N^\Phi - N^\Psi(1+N^\chi)
    (1+N^\Phi)\right],
  \label{Boltzmann}
\end{equation}
and similar equations for $\chi$ and $\Phi$. In the integrand we
denoted $N^\Psi(q,X)\to N^\Psi$, $N^\chi(\ell,X)\to N^\chi$ and
$N^\Phi(q-\ell,X)\to N^\Phi$. The possible values of $q_0$ are
$q_0=\pm \omega_\q$ (the dispersion relation).

We solve the equations in the relaxation time approximation. We
rewrite \eqref{JandG} using our condition that there is no net charge
in equilibrium
\begin{equation}
  Q^\Psi(t) = \pint4q 2q_0 \rh^\Psi(q) \delta N(q,t),
\end{equation}
where $\delta N$ is the deviation from equilibrium. For $\delta N$ the
linearized equation reads
\begin{equation}
  2q\d_X \delta N^\Psi = - h^2 \pint4\ell\rh^\chi\rh^\Phi \left[
    \delta N^\Psi\! \left(1+N^\chi+N^\Phi\right) + \delta N^\chi\!
    \left(N^\Psi-N^\Phi\right) +\delta N^\Phi\! \left(N^\Psi-
      N^\chi\right)\right].
  \label{linBoltzmann}
\end{equation}
We multiply this equation by $\rh^\Psi(q)$ and integrate over $q$. We
can realize that, because $\Phi$ is a real field,
\begin{equation}
  \c G_{rr}^\Phi(x,y)=\c G_{rr}^\Phi(y,x)\quad\Rightarrow\quad
  \c G_{rr}^\Phi(p,p') = \c G_{rr}^\Phi(-p',-p) \quad\Rightarrow\quad
  \delta \c W^\Phi(q,k) = \delta \c W^\Phi(-q,k).
\end{equation}
Since $\delta \c W = \rh^\Phi\delta N$ \cite{BuchFred}, this symmetry
renders the last term of \eqref{linBoltzmann} zero after integration.
We introduce
\begin{eqnarray}
  && \Gamma^\Psi(q) = \frac{h^2}{2q_0} \pint4\ell \rh^\chi(\ell)
  \rh^\Phi(q-\ell) (1+N(\ell_0)+N(q_0-\ell_0)) = \frac{-2\Im
    \Pi_R^\Psi(q)}{2q_0},\nn
  && \Gamma^\chi(q) = \frac{h^2}{2q_0} \pint4\ell \rh^\Psi(\ell)
  \rh^\Phi(q-\ell) (1+N(\ell_0)+N(q_0-\ell_0)) = \frac{-2\Im
    \Pi_R^\chi(q)}{2q_0},
  \label{GammaPsichi}
\end{eqnarray}
where $N(x)= (e^{\beta x}-1)^{-1}$ is the Bose--Einstein distribution,
and write
\begin{equation}
  \d_t Q^\Psi(t) = \pint4q \left[ - 2q_0 \delta N^\Psi(q,t) \rh^\Psi
    \Gamma^\Psi(q) + 2q_0 \delta N^\chi(q,t) \rh^\chi
    \Gamma^\chi(q)\right].
\end{equation}
The relaxation time approximation assumes that $\Gamma$ is only weakly
momentum-dependent, so we can factorize the integral, and we finally
obtain
\begin{eqnarray}
  && \d_t Q^\Psi(t) = - \Gamma^\Psi Q^\Psi(t) + \Gamma^\chi Q^\chi(t),
  \nn&& \d_t Q^\chi(t) = - \Gamma^\chi Q^\chi(t) + \Gamma^\Psi
  Q^\Psi(t).
\end{eqnarray}
This approximation satisfies the requirement that $Q^\Psi+Q^\chi=
\mathrm{const.}$, which is zero (the equilibrium value). Then the
solution of the above equations is
\begin{equation}
  Q^\Psi(t) = Q^\Psi(0) e^{-(\Gamma^\Psi+\Gamma^\chi) t},\qquad 
  Q^\chi(t) = -Q^\Psi(0) e^{-(\Gamma^\Psi+ \Gamma^\chi) t}.
\end{equation}
For $m_\Psi\gg m_\chi,m_\Phi$ we find
\begin{equation}
  \Gamma^\Psi = \frac{h^2}{16\pi m_\Psi} \left( 1+ 2N(\frac
    {m_\Psi}2)\right),\qquad
  \Gamma^\chi = \frac{h^2}{16\pi m_\chi} \beta M N(M)(1+N(M)),
\end{equation}
where $M=m_\Psi^2/2m_\chi$.

\section{Contributions coming from analytic defects}
\label{sec:cut}

The other important ingredient of the long-time evolution is the power
law tail coming from analytic defects of the spectral function.
These, in general (cf. Section~\ref{sec:longtime}), yield the time
dependence $A \cos(\Omega t+\phi)/t^{\alpha+1}$. The value of $\Omega$
and $\alpha$ depends on the theory, while the amplitude $A$ and the
phase $\phi$ come from the initial conditions.  In this section we
try to compute $\Omega$ and $\alpha$ for some special cases.

\subsection{Two-particle cut}

\begin{figure}[htbp]
  \begin{center}
    \includegraphics[height=2cm]{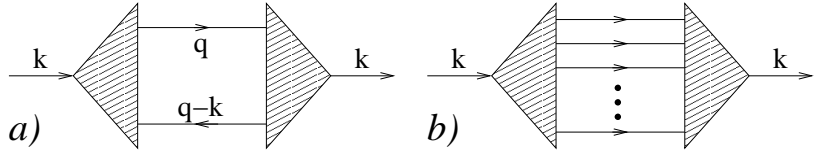}
    \caption{Two- and multiparticle intermediate state}
    \label{fig:twopartint}
  \end{center}
\end{figure}

Let us first consider a diagram where there is a two-particle
intermediate state, as is illustrated in Fig.~\ref{fig:twopartint}{\em
  a}.
We represent it as two momentum-dependent vertices denoted by
$\Gamma_1(q,k)$ and $\Gamma_2(q,k)$, their product being denoted by
$\Gamma=\Gamma_1\Gamma_2$. The intermediate particles have masses
$m_1$ and $m_2$, and we treat it as a cut diagram (with Keldysh
indices $21$ minus $12$). Its contribution then is
\begin{eqnarray}
  C_2(k) && = \pint4q \Gamma(q,k)\left[ G_1^{21}(q) G_2^{21}(k-q) -
  G_1^{12}(q) G_2^{12}(k-q)\right] \nn&&=
  4\pi^2 \pint4q \frac{\Gamma(q,k)}{4\omega_1\omega_2}[(1+N(q_0))
  (1+N(k_0-q_0))  - N(q_0)N(k_0-q_0)]\nn&&\hspace*{2cm}\times
  (\delta(q_0-\omega_1) - \delta(q_0+\omega_1))
  (\delta(k_0-q_0-\omega_2) - \delta(k_0 -q_0 +\omega_2)). 
  \label{2pthreshold}
\end{eqnarray}
The analytic defects of this function are thresholds, where the
phase-space, restricted by the constraints of the integration,
vanishes. There are 4 such points in this case: $k_0=\pm m_1\pm m_2$.
We restrict ourselves for positive $k_0$ and denote $k^{(1)}_0 =
m_1+m_2$ and $k^{(2)}_0=m_1-m_2$ (assuming $m_1\ge m_2$). At $k^{(1)}$
a cut starts, at $k^{(2)}$ it stops, so for the threshold behaviour we
choose $k_0 = k^{(1)}_0 + y$ or $k_0=k^{(2)}_0-y$. Since for $y\to 0$
the phase space vanishes ($\q\to 0$), we can expand everything to
second order with respect to $|\q|$.

In the $k^{(1)}$ case the relevant combination of the delta functions
of \eqref{2pthreshold} is
\begin{equation}
  \delta(q_0-\omega_1)\delta(k_0-q_0-\omega_2) \quad\Rightarrow\quad
  k_0 =\omega_1+\omega_2.
\end{equation}
Expanding $\omega_1+\omega_2$ around $k^{(1)}_0=m_1+m_2$ to second
order yields
\begin{equation}
  y =\left(\frac1{2m_1}+\frac1{2m_2}\right) \q^2,
  \label{quadexp}
\end{equation}
so that $|\q|$ vanishes as $y^{1/2}$. Assuming that $\Gamma$ does not
vanish in this limit (as the best case), and we substitute
$\omega_{1,2}\to m_{1,2},\, k_0=m_1+m_2$ and $\q=0$ where it is
possible, we find
\begin{equation}
  C \sim (1+N(m_1)+N(m_2)) \pint3{\q} \delta\left( y -
  \left(\frac1{2m_1}+\frac1{2m_2}\right) \q^2\right) \sim y^{1/2},
\end{equation}
which can be seen after performing the integral. This is the same as
the zero-temperature result \cite{ELOP}, only the amplitude is
modified by $(1+N(m_1)+N(m_2))$.

The $k^{(2)}$ case goes along the same lines. We choose the
combination
\begin{equation}
  \delta(q_0-\omega_1)\delta(k_0-q_0+\omega_2) \quad\Rightarrow\quad
  k_0 =\omega_1-\omega_2 \quad\Rightarrow\quad y = \left(\frac1{2m_2}
  -\frac1{2m_1} \right) \q^2,
\end{equation}
and therefore
\begin{equation}
  C\sim (N(m_1)-N(m_2)) y^{1/2}.
\end{equation}
This is a finite temperature effect, the threshold of the Landau cut
\cite{LeBellac}.

For our concrete example described by the Lagrangian \eqref{Lag}, only
equal-mass two-particle intermediate states are allowed. In that case
the first threshold is just a zero temperature one, starting at $2m_i$
where $m_i\in\{m,M,m_S\}$. The second case, however, is not present,
since the delta constraints then give $\delta(k_0)$. So there the
important cut contribution has $\Omega=2m_i$ and $\alpha=1/2$. That is
we obtain an oscillating contribution which has presumably negligible
effect on real physical processes.

\subsection{Four-particle intermediate state}

Four-particle intermediate states may contain cuts, and this will be
analyzed in the next subsection. Here we concentrate on other possible
analytic defects.

On the same footing as before we can write up the contribution coming
from the four-particle intermediate state
\begin{equation}
  C_4(k)=\pint4{q_1}..\frac{d^4q_4}{(2\pi)^4}\,\delta(k-{\textstyle\sum}
  q) \Gamma(k,\{q\}) \left[(1+N_1)..(1+N_4)-N_1..N_4\right]
  \rh_1(q_1).. \rh_4(q_4),
\end{equation}
where $N_i=N(q_{0i})$. We change $q\to -q$ in the first term and write
$C_4(k)= C_4^+(k) - C_4^-(k)$ where
\begin{equation}
  C_4^{\pm}(k) =\pint4{q_1}..\frac{d^4q_4}{(2\pi)^4}\,
  \delta(k\pm{\textstyle\sum} q) \Gamma(k,\{\mp q\}) N_1..N_4\, \rh_1
  .. \rh_4.
\end{equation}
In fact, this is a general form; we can write also for the
two-particle contribution \eqref{2pthreshold} $C_2(k)= C_2^+(k) -
C_2^-(k)$, where
\begin{equation}
  C_2^{\pm}(k) =\pint4{q_1}\frac{d^4q_2}{(2\pi)^4}\,
  \delta(k\pm{\textstyle\sum} q) \Gamma(k,\{\mp q\}) N_1N_2\,
  \rh_1\rh_2.
\end{equation}

We now make a simplification, and assume that for some reason $\Gamma$
is factorized as $\Gamma(k,\{q\}) = \Gamma^{(1,2)}(k,q_1,q_2)
\Gamma^{(3,4)}(k,q_3,q_4)$. This means a loss in generality; but we
hope that the main consequences still remain true. We introduce an
auxiliary variable $Q$ and write
\begin{eqnarray}
  C_4^+(k)= \pint4Q && \pint4{q_1}\!\frac{d^4q_2}{(2\pi)^4}\,
  \delta(Q+q_1+q_2)\Gamma^{(1,2)}(k,-q_1,-q_2) N_1 N_2 \rh_1(q_1)
  \rh_2(q_2) \nn\times && \pint4{q_3}\!\frac{d^4q_4}{(2\pi)^4}\,
  \delta(k-Q+q_3+q_4)\Gamma^{(3,4)}(k,-q_3,-q_4) N_3
  N_4\rh_3(q_3)\rh_4(q_4).
\end{eqnarray}
On the right-hand side we recover the convolution of two $C_2^+$
contributions; in real time they are simply multiplied. The complete
four-particle contribution can be written as
\begin{equation}
  C_4(t) = C_2^+(t) {C'_2}^+(t) - C_2^-(t) {C'_2}^-(t).
  \label{C2prod}
\end{equation}
At zero temperature $N(q_0)\to-\Theta(-q_0)$, thus $C_2^+$ contains
only the positive energy cuts and $C_2^+ {C'}_2^+$ oscillates with
frequency $\Omega+\Omega'$. At finite temperature, however, the
negative energy cut appears as well, and, for the same threshold
values, we have
\begin{equation}
  \frac{A_- e^{i\Omega t} + A_+ e^{-i\Omega t}}{t^{3/2}} 
  \frac{A'_- e^{i\Omega t} + A'_+ e^{-i\Omega t}}{t^{3/2}} =
  \frac{A_- A'_+ + A_+ A'_-}{t^3} + \mathrm{oscillating}.
\end{equation}

Diagrammatically it comes from the diagram where two lines are
incoming, two are outgoing, or vice versa. Kinematically, with the
external line, it is a $2\to3$ or $3\to2$ process. Also from here we
see that this contribution must be absent at zero temperature. We
can assess the temperature dependence by the typical factor for 2
incoming and 2 outgoing internal lines,
\begin{equation}
  (1+N(m_1))(1+N(m_2)) N(m_3) N(m_4).
\end{equation}
For low temperatures the contribution is therefore
Boltzmann-suppressed.

We can also have cancellations at finite temperature, if the $+$ and
$-$ parts in \eqref{C2prod} cancel each other. $C^+$ and $C^-$ are
related as (writing out explicitly the dependence on the $\Gamma$
vertex function)
\begin{equation}
  C^+(\Gamma,-k) = C^-(\tilde \Gamma,k),\quad\mathrm{where}\quad
  \tilde\Gamma(k,\{q\}) = \Gamma(-k,\{-q\}).
\end{equation}
Cancellation occurs, if $\tilde\Gamma= \pm\Gamma$ for small $k$. For
generic $F$ at finite temperature we do not expect such a symmetry,
therefore a net effect will remain.

\begin{figure}[htbp]
  \centering
  \includegraphics[height=2.5cm]{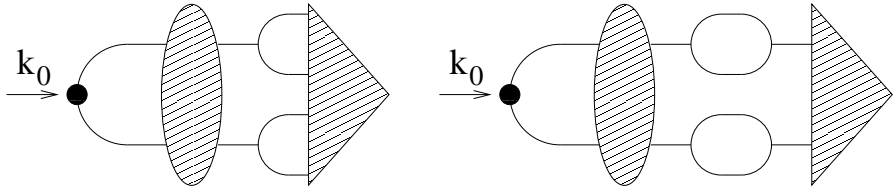}
  \caption{The realization of a four-particle internal state. The shaded
    oval denotes ladder resummation.}
  \label{fig:c4}
\end{figure}

How can this phenomenon show up in a diagram? If we calculate $Q(k_0)$
for small $k_0$, we encounter ladder diagrams, as is shown in
Fig.~\ref{fig:fermdiag}. The ending of the ladder (the triangle at the
right-hand side) contributes to the numerator (the constant $C$ in
eq.~\eqref{Glongtime}). Here can sit the four-particle intermediate
state of this section, see for example Fig.~\ref{fig:c4}.
The $k_0\approx0$ behaviour of the charge is therefore
\begin{equation}
  Q(k_0\to0)\sim \frac{\zeta_2 + \# C_4(k_0) h^4 \zeta_2 +
  \# C_4(k_0) h^2 \zeta_4}{k_0+i\Gamma}.
\end{equation}
The pole and the $k_0=0$ defects are therefore separated and do
not affect the behaviour of each other. When computing the numerical
coefficient of the $k_0=0$ defect we substitute $k_0=0$ in the
numerator, which means an enhancement factor of $1/\Gamma$.

Four-particle initial conditions can also come from the quadratic
response (as a product of two two-particle operators). Having the same
effect, it is natural to assume that their contribution is also in the
same order of magnitude, i.e. $\zeta_4\sim \zeta_2^2$. Since $\zeta
\ll 1$ (linear response regime) we find suppression in this
term\footnote{This hierarchy, however, does not necessarily hold: wild
  initial conditions can violate it, then the $1/t^3$ time dependence
  also has a coefficient of $\c O(1)$.}. To have only one small
parameter in the system we will assume that $\zeta_2\sim h^2$.

Summarizing, the $1/t^3$ contribution of the four-particle
intermediate state is suppressed by several effects compared to the
Boltzmann pole. It has a perturbative suppression of $h^2$; it is a
finite-temperature effect, so that it is Boltzmann-suppressed at small
temperatures; it has two contributions with opposite sign; and finally
it is a multiloop effect, which probably means also a numerical
suppression.

\subsection{Multiparticle cuts and other defects}

The determination of the threshold behaviour described in the case of
two-particle intermediate states can be generalized to multiparticle
intermediate states (cf. Fig.~\ref{fig:twopartint}{\em b}). We expect
thresholds when the external $k_0$ plus the $N$ internal particles
represent a decay process ($1\to N$ or $N\to 1$), since scattering
processes do not have cuts. There always exists a possibility, when
all of the internal particles are outgoing, so that $k^{(1)}_0=\sum_i
m_i$ is the threshold value; this is also the zero-temperature case.
At finite temperature an additional threshold is possible (for
positive $k_0$), when one particle is outgoing, the rest are incoming.
Then we obtain a threshold at $k_0^{(2)} = m_1 - \sum_{i=2}^N m_i$, if
it is positive.

Near the threshold, in the massive case, we can perform the same
expansion as in \eqref{quadexp}, and also there is a
momentum-conserving delta function $\delta(\sum \q_i)$. Therefore what
remains is
\begin{equation}
  \int d^3\q_1\dots d^3\q_N \delta\left( y - \sum_i
  \frac{\q_i^2}{2m_i}\right) \delta\left( \sum_i q_i \right) \sim
  y^{\frac{3N}2} y^{-1} y^{-\frac32} \sim y^{\frac{3(N-1)}2-1}.
\end{equation}
The possible values for $\Omega$ thus are $\sum_i m_i$ and
$m_1-\sum_{i>1} m_i$ (if it is positive), in both cases $\alpha=
\frac{3(N-1)}2-1$ (which is also the zero-temperature result).

Similarly to the two times two-particle cut case, multiparticle cuts
can also be multiplied in real space to generate new defects. This
means, however, a faster decay than $1/t^3$.

\section{Conclusions}
\label{sec:conclusion}

In this paper we tried to describe the time evolution of the charge
operator $Q$ in a complex scalar system, starting from an initial
state not too far from thermal equilibrium. The initial state was
prepared using an external action that modifies the time evolution
for a given period. We applied linear response theory in the external
action to describe time evolution. As a computing method we used real
time perturbation theory.

The long-time behaviour of $Q$ is determined by the $QF$ spectral
function in the linear response theory, where $F$ is an arbitrary
operator ($[Q,F]\neq0$). If $Q$ were conserved, there would be
$\rh^{(QF)}(k_0) \sim \delta(k_0)$. For the non-conserving case two
types of contributions can be identified (cf. \cite{JMY,Boyan,JPSZ}).
One is coming from the broadening of the $\delta(k_0)$ pole of the
conserved $Q$-case, yielding exponential damping for long times. In
perturbation theory it manifests itself as pinch singularities in the
ladder diagrams, giving $1/k_0^n$ type of IR divergences at any order
in the coupling. Resummation of these singularities results in
linearized Kadanoff--Baym equations. Working in real space,
$1/k_0^{n}$ singularities correspond to the secular terms $t^{n-1}$,
and their resummation also leads to quantum Boltzmann equations
\cite{Boyan-DRG}.

Analytic defects in $\rh^{(QF)}(k_0)$ also contribute to long-time
physics: it gives an $e^{-i\Omega t}/t^{\alpha+1}$ oscillating
power-law decay, where $\Omega$ is the position of the defect,
$\alpha$ is the index of $\rh$ near to the defect. For multiparticle
cuts we have $\Omega=\sum_i m_i$ or $m_i-\sum_{i>1}m_i$ (if it is
positive: a multiparticle Landau damping), and index
$\alpha=3(N-1)/2-1$, where $N$ is the number of intermediate particles
and $m_i$ are their masses.  These contributions are oscillating with
the particle mass, so that they can hardly influence physics on IR
scales.  At finite temperature, however, these cuts can be combined,
and from their superposition the oscillating behaviour can drop out.
The lowest order diagram giving this result contains 4 intermediate
particles and yields $1/t^3$ type of decay.

So finally, starting from a generic initial state characterized by
$F$, we expect that the time evolution of the charge for large times
follows
\begin{equation}
  Q(t) = Q_1 e^{-\Gamma t} + \frac{Q_2}{t^3} + \mathrm{oscillating}.
\end{equation}
The relative weights of the Boltzmann pole and the cut contributions
contain free parameters, depending on the initial state, and they
contain perturbative factors. The pole is present already at the free
level, we thus expect $Q_1 = \c O(1)\,+ $ perturbative corrections.
The amplitude of $Q_2$ is reduced by several effects; for example it
is suppressed by $h^2$ and by the Boltzmann factor for small
temperatures. Thus, for a generic situation we expect that exponential
damping dominates the time evolution for intermediate times and only
after a long time will the power law take over (for a numerical
simulation example see, \cite{BorsSzep}). Still, because of its slow
decay, it can be relevant in the explanation of small deviations from
equilibrium (such as baryogenesis). This should be examined in the
future.

Another task for the future is to include non-linear effects and see
how the full Boltzmann equation gets modified due to the presence of
cut contributions.

\section*{Acknowledgment}

The author would like to thank to M. Laine for his help and advice.
He also readily acknowledges the useful discussions with D.~Boyanovsky,
D.~B\"odeker, W.~Buchm\"uller, Z.~Fodor, A.~Patk\'os and
M.~Pl\"umacher. This work was partially supported by the Hungarian
Science Fund (OTKA).

\appendix

\section{Appendix: pinch singularities}
\label{sec:pinch}

The pole structure of the retarded and advanced propagators is
represented by
\begin{equation}
  D_{R/A}(q) = \frac1{(q_0\pm i\ep)^2-\omega^2} =
  \frac1{2\omega}\left[ \frac1{q_0-\omega\pm i\ep}
    -\frac1{q_0+\omega \pm i\ep}\right]
  \label{DRAform}
\end{equation}
where $\omega(\q)$ is the dispersion relation. They have poles at
$\pm\omega -i\ep$ for the retarded Green's function and $\pm\omega
+i\ep$ for the advanced one. Their product thus has poles at
$\pm\omega\pm i\ep$, which pinch the real axis for $\ep\to0$,
resulting in double poles which cannot be regularized. We do not
expect such problems for products like $D_RD_R$ or $D_AD_A$.

In our case the momenta of the retarded and advanced propagators are
offset by $k$.  For finite $k$ we do not expect singularities, but the
$k\to 0$ limit must be divergent.

First we examine the $D_R(q+k)D_A(q)$ product for small $k=(k_0,0)$.
We will use the identity
\begin{equation}
  \frac1{x+i\ep}= \cP\left(\frac1x\right) -i\pi \delta(x)
\end{equation}
where ${\cal P}$ means principal value, so that
\begin{equation}
  D_{R/A}(q) = \frac1{2\omega}\left[\cP\left(\frac1{q_0-\omega}\right)
    \mp i\pi\delta(q_0-\omega) - \{\omega\to -\omega\}\right].
\end{equation}
For the product we obtain
\begin{eqnarray}
  D_R(q+\frac k2)D_A(q-\frac k2) = \frac1{4\omega^2} \biggl[
  \!&&i\pi\cP \left(\frac1{k_0}\right) \left(\delta(q_0-\frac{k_0}2
  -\omega) + \delta(q_0+\frac{k_0}2-\omega)\right) \nn&&+
  \pi^2\delta(k_0)\delta(q_0-\omega) +\,
  \{\omega\to-\omega\}\biggr] +\textrm{non-singular},
  \label{productofGRGA}
\end{eqnarray}
the non-singular terms having finite limit at $k\to 0$. Rewriting this
expression we obtain
\begin{equation}
  D_R(q+\frac k2)D_A(q-\frac k2) = i\cP\left( \frac1 {2q_0 k_0}\right)
  \rh(q) + o \left(\frac1{k_0}\right),
  \label{pinchGRGA}
\end{equation}
where $\rh$ is the free spectral function; $o(1/x)$ means that
$\lim_{x\to0}(x\,o(1/x))=0$, which is trivial for non-singular terms,
and true for the delta term since $x\delta(x)=0$.

In a real model we can write for the propagator $G(q) = d(q) D(q)$,
where $d$ is the Klein--Gordon divisor\footnote{For fields with
  multiple mass shell we obtain a sum of these terms.}
\cite{LandvanWeert}.

We can obtain \eqref{pinchGRGA} in a different way, using
$i(G_R-G_A)=\rh$ and that $G_RG_R$ is non-singular. We write
\begin{equation}
  G_R(q+\frac k2)G_A(q - \frac k2) = G_R(q+\frac k2)\,i\rh(q-\frac k2)
  - G_R(q+k)G_R(q).
\end{equation}
The second term is finite for $k\to0$, while in the first term at
$k=0$ we put a propagator on mass shell, which is clearly divergent.
The numerator of $G_R(q+k/2)$ for small $k$ reads
\begin{equation}
  q_0^2+q_0k_0+\frac{k_0^2}4-\omega_{\q+\k/2}^2 \to 2q_0k_0
  +\omega_{\q-\k/2}^2-\omega_{\q+\k/2}^2 \to 2q_0k_0 -
  2\k\cdot\d\omega,
\end{equation}
where we have power-expanded $\omega_{\q\pm\k/2}$. Therefore
\begin{equation}
  G_R(q+\frac k2)G_A(q-\frac k2) = \frac{d(q)}{2q_0k_0 -
  2\k\cdot\d\omega}\,i\rh(q) + \dots,
\end{equation}
where $d(q)$ is the Klein--Gordon divisor. It is also clear that any
other $G_R$ factor yields the same $d(q)/(2q_0k_0 - 2\k\cdot\d\omega)$
contribution. Additional $G_A$ factors yield $-d(q)/(2q_0k_0
-2\k\cdot\d\omega)$ contribution, as can be seen by using the above
formulae for $G_R=-i\rh+G_A$.

In the paper we use $k=(k_0,0)$ momentum, but all the formulae can
easily be generalized to finite $\k$ simply by $2q_0k_0\to 2q_0k_0 -
2\k\cdot\d\omega$.

Pinch singularities occur in other products as well. In the R/A
formalism \cite{RAformalism} the propagators are $G_R,\, G_A$ and
$G_{rr} = (G_{12}+G_{21})/2$. Since $G_{rr}\sim\rh $ also puts the
propagation on the mass shell, we find
\begin{equation}
  G_{R/A}(q+\frac k2) G_{rr} (q-\frac k2) = \frac{d(q)}{2q_0k_0}
  G_{rr}(q).
\end{equation}
On the other hand $G_{rr}(q+k)\cdot G_{rr}(q)\sim \delta(k_0)$,
therefore it yields an $o(1/k_0)$ contribution. So in the two-particle
intermediate states the possible pinch enhancements are coming from
$G_RG_A$ and $G_{rr} G_{R/A}$ only.

\end{document}